\pdfoutput=1
\documentclass[a4paper,11pt]{article}
\usepackage{amsmath}
\usepackage{amssymb}
\usepackage{amsfonts}
\usepackage{graphicx}
\usepackage{animate}

\usepackage[utf8]{inputenc} 

\usepackage[square,numbers,comma,sort&compress]{natbib}
\usepackage[colorlinks=true,citecolor=blue,linkcolor=purple,urlcolor=purple]{hyperref}
\usepackage{geometry}
\usepackage[usenames,dvipsnames,svgnames,table]{xcolor}
\usepackage{booktabs}
\usepackage{slashed}

\def\varabstract{ }
\def\varkeywords{ }
\def\vararxivnumber{ }
\def\vartitle{ }
\def\varsubtitle{ }
\renewcommand{\title}[1]{\gdef\vartitle{#1}}

\renewcommand{\abstract}[1]{\gdef\varabstract{#1}}
\newcommand{\keywords}[1]{\gdef\varkeywords{#1}}

\newtoks\authtoks
\renewcommand{\author}[2][]{%
	\authtoks=\expandafter{\the\authtoks#2$^{#1}$\ }%
}
\newtoks\affiltoks
\newcommand{\affiliation}[2][]{%
    \affiltoks=\expandafter{\the\affiltoks{\item[$^{#1}$]#2}}%
}
\newtoks\emailtoks\newcounter{emailcounter}%
\newcommand{\emailAdd}[1]{%
\ifnum\theemailcounter>0\emailtoks=\expandafter{\the\emailtoks, \typeemail{#1}}%
\else\emailtoks=\expandafter{\typeemail{#1}}%
\fi
\stepcounter{emailcounter}}
\newcommand{\typeemail}[1]{\href{mailto:#1}{\tt #1}}

\renewcommand\maketitle{
	\newgeometry{margin=2cm}
	\pagestyle{empty}\setcounter{page}{0}
	{\huge\flushleft\sffamily\bfseries\vartitle\\\Large\varsubtitle\par}
\vskip6ex
{\large\bfseries\raggedright\sffamily\the\authtoks\par}
\vskip2ex
\begin{list}{}{%
\setlength{\leftmargin}{0.28cm}%
\setlength{\labelsep}{0pt}%
\setlength{\itemsep}{-3pt}%
\setlength{\topsep}{-\parskip}}
\itshape\small%
\the\affiltoks
\end{list}
\vskip2ex
\noindent\hspace{0.28cm}\begin{minipage}[l]{.9\textwidth}
\begin{flushleft}
\textit{E-mail:} \the\emailtoks
\end{flushleft}
\end{minipage}
\vskip5ex
\noindent{\renewcommand\baselinestretch{.9}\textsc{Abstract:}}\ \varabstract
\vskip5ex 
\if!\varkeywords!\else\noindent{\textsc{Keywords:}}\ \varkeywords \vskip2ex\fi
\if!\vararxivnumber!\else\noindent{\textsc{ArXiv ePrint:}} \href{http://arxiv.org/abs/\vararxivnumber}{\vararxivnumber}\vskip2ex\fi

\newpage
\restoregeometry
\pagestyle{plain}

\setcounter{footnote}{0}
} 

\usepackage[square,numbers,comma,sort&compress]{natbib}

\definecolor{MS}{rgb}{0,0,1}

\usepackage{ifthen}

	\newcommand{\barlimc}[7]{
  \pgfmathparse{\mypos+0.3}
  \edef\mypos{\pgfmathresult}
		\node[left,scale=0.6] at (0,\mypos) {#1};
		\pgfmathparse{#3 > 5 ? 1 : 0}
		\ifthenelse{\pgfmathresult=1}{
			\fill[#2] ($(0,\mypos)+(0,-0.1)$) rectangle +(5,0.2);
			\fill[white] ($(0,\mypos)+(3.5,-0.1)$) rectangle +(0.3,0.2);
			\draw[decoration={zigzag},decorate,#2,very thick] (3.4,\mypos) to +(0.5,0);
			\node[left,scale=0.6] at (5,\mypos) {#3};
			}{
			\fill[#2] ($(0,\mypos)+(0,-0.1)$) rectangle +(#3,0.2);
			\node[left,scale=0.6] at (#3,\mypos) {#3};
		}		
		\fill[#4] ($(0,\mypos)+(0,-0.1)$) rectangle +(#5,0.2);
		\node[left,scale=0.6] at (#5,\mypos) {#5};
		\fill[#6] ($(0,\mypos)+(0,-0.1)$) rectangle +(#7,0.2);
		\pgfmathparse{#7 <0.3 ? 1 : 0}
		\ifthenelse{\pgfmathresult=1}{
			\node[right,scale=0.6] at (0,\mypos) {#7};
		}{
		\node[left,scale=0.6] at (#7,\mypos) {#7};
	}
}

\title{Color-octet scalar decays to a gluon and an electroweak gauge boson in the Manohar-Wise model}

\author[1]{Alper Hayreter,}\emailAdd{alper.hayreter@ozyegin.edu.tr}
\author[2]{and German Valencia}\emailAdd{german.valencia@monash.edu}

\affiliation[1]{Department of Natural and Mathematical Sciences, Ozyegin University, 34794 Istanbul Turkey.}

\affiliation[2]{School of Physics and Astronomy, Monash University,
Wellington Road, Clayton, VIC-3800, Australia}

\abstract{We present one loop results for the amplitudes giving rise to couplings between a color octet scalar, a gluon, and an electroweak gauge boson. These amplitudes could signal new physics  in $\gamma$ jet, $Z$ jet and  $W$  jet   production at the LHC. We compute the relevant branching ratios and identify regions of parameter space where these decay modes become important. This can happen for scalar masses below the threshold for decay into heavy quark pairs ($t\bar t$ and $t\bar b$); or for small Yukawa couplings in which case  the colored scalars are fermiophobic. In the case of light scalars, ${\cal B}(S\to \gamma g)$  can reach up to 10\% whereas  ${\cal B}(S\to Z g)$  can reach a few percent. In the fermiophobic region of parameter space, ${\cal B}(S\to \gamma g)$ and ${\cal B}(S\to Z g)$ can reach up to 72\% and 28\% respectively, whereas ${\cal B}(S\to g g)$ can be 100\%. For the charged scalar, the decay mode ${\cal B}(S^\pm \to W^\pm g)$ can become dominant  in both scenarios.}


\begin{document}

\maketitle

\section{Introduction}

Many extensions of the standard model (SM) contain colored scalars that give rise to rich phenomenology. These include the sgluons and squarks  of supersymmetry and scalar leptoquarks, for example. Multi-Higgs models can also include scalar representations charged under the color group. One compelling example is the 
Manohar-Wise model (MW) \cite{Manohar:2006ga} in which scalars transforming as a color octet, electroweak doublet are introduced. This particular representation can couple directly to quarks while respecting minimal flavor violation, thus naturally satisfying constraints from flavor physics.

The MW  model contains fourteen new parameters in the scalar potential and an additional four parameters in the Yukawa sector and has been studied at length in the literature. For example, the new scalars can modify the $Hgg$ coupling at one-loop and alter significantly the Higgs production and decay phenomenology \cite{Manohar:2006ga,He:2011ti,Dobrescu:2011aa,Bai:2011aa,Cacciapaglia:2012wb,Cao:2013wqa,He:2013tia}. The new scalars also affect precision electroweak measurements \cite{Manohar:2006ga,Gresham:2007ri,Burgess:2009wm} and flavor physics \cite{Grinstein:2011dz,Cheng:2015lsa,Martinez:2016fyd,Faber:2018afz} and all this leads to constraints on its parameters. In addition to these phenomenological constraints, the parameter space is restricted by theoretical considerations such as unitarity and vacuum stability \cite{Reece:2012gi,He:2013tla,Cheng:2016tlc,Cheng:2017tbn}. After taking these constraints into account the  MW model can still produce many observable effects at the  LHC \cite{Gerbush:2007fe,Arnold:2011ra,He:2011ws,Kribs:2012kz,Hayreter:2017wra}.

In this paper we study decay modes of the MW scalars that have not received much attention thus far, namely the one-loop induced processes connecting a scalar to a gluon and an electroweak gauge boson $W^\pm, Z$ or $\gamma$. The MW scalars can be pair produced at tree-level through their QCD couplings and subsequently decay. Most of the time they will decay into pairs of heavy quarks through their Yukawa couplings. The effective couplings we compute in this paper induce decays into electroweak gauge bosons and jets that typically occur at much lower rates.  However, there are  regions of parameter space where these decay modes  become dominant. 

The phenomenology of new physics in $\gamma j$ and $Z j$ final states at LHC has received some attention in the literature. It has been recently considered in the context of a pseudoscalar color octet $\pi_8$ \cite{Belyaev:2016ftv}, where it is argued that $\gamma j$, in particular, is a clean channel due to the presence of an energetic photon. Ref.~\cite{Carpenter:2015gua,Carpenter:2020hyz} have studied the decays of sgluons into $\gamma j$ via squark loops.  Early studies of an apparent di-jet anomaly reported by CDF \cite{Aaltonen:2011mk} considered  tree-level processes with color octet scalars resulting in $\gamma j,~Z j,~{\rm or~}W^\pm j$ final states \cite{Carpenter:2011yj,Enkhbat:2011qz}. Very recently there has also been a phenomenological study presenting constraints on $\gamma j$ final states at LHC  \cite{Cacciapaglia:2020vyf}. 

This paper is organized as follows. In section~2 we review the relevant features of the MW model paying particular attention to the sector of the model that is relevant for this study. In section~3 we present explicit one-loop results for the $SVg$ vertices including quark and scalar loop contributions. In section~4 we discuss the regions of parameter space where these decay modes can become important. In section~5 we present numerical results for benchmark points illustrating the branching ratios ${\cal B}(S\to Vg)$ that can be reached.  A phenomenological study of  signals for these modes at LHC is beyond the scope of this paper, but we provide preliminary comments by comparing our one-loop vertices with the recent study of \cite{Cacciapaglia:2020vyf}.

\section{The Model}

In the MW model, the new scalar field $S$ transforms as $(8,2,1/2)$ under the SM gauge group $SU(3)_C\times SU(2)_L\times U(1)_Y$. Numerous new couplings appear in the scalar potential and in the Yukawa sector. The possible Yukawa couplings reduce to two complex numbers once minimal flavor violation is imposed \cite{Manohar:2006ga}, 
\begin{equation}
{\cal L}_Y=-\eta_U e^{i\alpha_{U}} g^{U}_{ij}\bar{u}_{Ri}T^A Q_j S^A - \eta_D e^{i\alpha_{D}}g^{D}_{ij}\bar{d}_{Ri}T^A Q_j S^{\dagger A} + h.c.
\label{yukc}
\end{equation}
Here $Q_i$ are the usual left-handed quark doublets and $S^a$ are the new scalars written as $S=S^aT^a$ $SU(3)$ generators normalized as ${\rm Tr} (T^aT^b)=\delta^{ab}/2$. The  matrices $g^{U,D}_{ij}$ are the same as the  Higgs couplings to quarks, and the overall strength of the interactions is given by $\eta_{U,D}$ along with their phases $\alpha_{U,D}$. The latter introduce Charge-Parity (CP) violation beyond the SM and contribute for example to the Electric Dipole Moment (EDM) and Chromo-Electric Dipole Moment (CEDM) of quarks \cite{Manohar:2006ga,Burgess:2009wm,He:2011ws,Martinez:2016fyd}. 

The most general renormalizable scalar potential is given in Ref.~\cite{Manohar:2006ga} The new couplings we derive in this work will only depend on the following terms, 
\begin{eqnarray}
&&V=\lambda\left(H^{\dagger i}H_i-\frac{v^2}{2}\right)^2+2m_s^2\ {\rm Tr}S^{\dagger i}S_i +\lambda_1\ H^{\dagger i}H_i\  {\rm Tr}S^{\dagger j}S_j +\lambda_2\ H^{\dagger i}H_j\  {\rm Tr}S^{\dagger j}S_i 
\nonumber \\
&&+\left( \lambda_3\ H^{\dagger i}H^{\dagger j}\  {\rm Tr}S_ iS_j +\lambda_4\ e^{i\phi_4}\ H^{\dagger i} {\rm Tr}S^{\dagger j}S_ jS_i +
\lambda_5\ e^{i\phi_5}\ H^{\dagger i} {\rm Tr}S^{\dagger j}S_ iS_j 
+{\rm ~h.c.}\right)
\label{potential}
\end{eqnarray}
where $v\sim 246$~GeV. 
The number of parameters in Eq.(\ref{potential}) can be further reduced by theoretical considerations: first $\lambda_3$ can be chosen to be real by a suitable definition of $S$; custodial $SU(2)$ symmetry implies the relations $2\lambda_3=\lambda_2$ (and hence $m_{S^+}=m_I$) \cite{Manohar:2006ga} and $\lambda_4=\lambda_5^\star$ \cite{Burgess:2009wm}; and   CP conservation removes all the phases, $\alpha_U$, $\alpha_D$, $\phi_4$ and  $\phi_5$. After symmetry breaking,  the Higgs vev in Eq.(\ref{potential})  splits  the octet scalar masses as,
\begin{eqnarray}
m^2_{S^{\pm}} =  m^2_S + \lambda_1 \frac{v^2}{4},&&
m^2_{S_{R,I}} =  m^2_S + \left(\lambda_1 + \lambda_2 \pm 2 \lambda_3 \right) \frac{v^2}{4}. 
\end{eqnarray}
In our calculation, the parameters $\lambda_{1,2,3}$  simply control this mass splitting and will be traded for the scalar masses. The triple scalar coupling depends on $\lambda_{4,5}$, and it determines the magnitude of the scalar loop contributions to the $Sgg$, $Sg\gamma$, $SgZ$ and $S^\pm gW^\pm$ we compute next. Finally, $\eta_{U,D}$ control respectively the strength of the $Stt$ and $Sbb$ interactions. 

The effective one-loop  couplings of the form $SVg$ can be written in terms of two (dual) field strength tensors $F(\tilde F)_{R,I}^{Vg}$ as,
\begin{align}
{\cal L}_{Sg g }&=\frac{\alpha_s}{8\pi v}\Bigg[\left(F_R^{gg}\ G_{\mu\nu}^A G^{B\mu\nu} + \tilde F_R^{gg}\ \tilde{G}_{\mu\nu}^A G^{B\mu\nu}\right) S_R^C +\left(F_I^{gg}\ G_{\mu\nu}^A G^{B\mu\nu} +\tilde F_I^{gg}\ \tilde{G}_{\mu\nu}^A G^{B\mu\nu}\right) S_I^C\Bigg] d^{ABC},\nonumber\\ \nonumber\\
{\cal L}_{S\gamma g}&=\frac{\sqrt{\alpha\alpha_s}}{3\pi v}\Bigg[\left(F_R^{\gamma g}\ G_{\mu\nu}^A A^{\mu\nu} + \tilde F_R^{\gamma g}\ \tilde{G}_{\mu\nu}^A A^{\mu\nu}\right) S_R^B\ +\left(F_I^{\gamma g}\ G_{\mu\nu}^A A^{\mu\nu} + \tilde F_I^{\gamma g}\ \tilde{G}_{\mu\nu}^A A^{\mu\nu}\right) S_I^B\Bigg] \delta^{AB},\nonumber\\ \nonumber\\
{\cal L}_{SZg}&=\frac{\sqrt{\alpha\alpha_s}}{12\pi v}\Bigg[\left(F_R^{Z g}\ G_{\mu\nu}^A Z^{\mu\nu} +\tilde F_R^{Z g}\ \tilde{G}_{\mu\nu}^A Z^{\mu\nu}\right) S_R^B\ +\left(F_I^{Z g}\ G_{\mu\nu}^A Z^{\mu\nu} + \tilde F_I^{Z g}\ \tilde{G}_{\mu\nu}^A Z^{\mu\nu}\right) S_I^B\Bigg] \delta^{AB},\nonumber\\ \nonumber\\
{\cal L}_{SWg}&=\frac{\sqrt{\alpha\alpha_s}}{12\pi v}\left(F^{W g}\ G_{\mu\nu}^A W^{\pm\mu\nu} + \tilde F^{W g}\ \tilde{G}_{\mu\nu}^A W^{\pm\mu\nu}\right)S^{\mp B}\ \delta^{AB}
\label{defff}
\end{align}
where $G^A_{\mu\nu}$, $A_{\mu\nu}$,  $Z_{\mu\nu}$ and  $W_{\mu\nu}$ are the gluon, photon, $Z$ and $W$ field strength tensors respectively and $\tilde G^{A\mu\nu} = (1/2)\epsilon^{\mu\nu\alpha\beta}G^A_{\alpha\beta}$.

Explicit one-loop results for these factors in the MW model are presented in the next section. Of these couplings, only ${\cal L}_{Sgg}$ exists in the literature and we find a sign difference with that result that we describe below. These effective vertices receive their main contributions from top-quark and color-octet scalar loops. The bottom-quark loop is important only for regions of parameter space where $|\eta_D|>>|\eta_U|$.  

\section{Explicit one-loop results in the MW model}

We perform the calculation with the aid of a number of software packages. We first implement the model in FeynRules \cite{Christensen:2009jx,Alloul:2013bka} to generate  FeynArts \cite{Hahn:2000kx} output where LoopTools \cite{Hahn:1998yk} is used to compute the one loop diagrams and simplification is assisted with FeynCalc \cite{Mertig:1990an,Shtabovenko:2016sxi}, FeynHelpers \cite{Shtabovenko:2016whf} and Package-X \cite{Patel:2015tea}. We present our result without assuming custodial or CP symmetries. 

\subsection{\boldmath\underline{$S_{R,I}\to g\ g$}}

\begin{figure}[ht]
\centering
\includegraphics[scale=0.45]{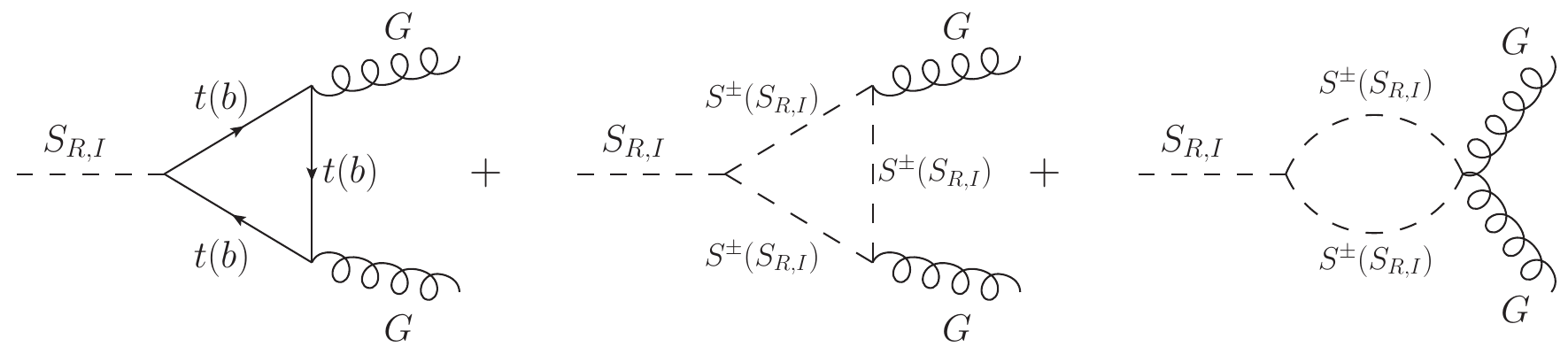}
\caption{One-loop diagrams contributing to the factors appearing in Eq.~(\ref{defff}) for the  $S_{R,I}gg$ coupling.}
\label{f:sgg}
\end{figure}
The diagrams responsible for the $Sgg$ couplings are shown in Figure~\ref{f:sgg} and result in form factors  given by
\begin{align}
F_R^{gg} &=\Bigg\{
\eta_U c_U\ I_q\left(\dfrac{m_t^2}{m_R^2}\right)
+\eta_Dc_D\ I_q\left(\dfrac{m_b^2}{m_R^2}\right)\nonumber \\
&+\dfrac{9}{4}\dfrac{v^2}{m_R^2}\ (\lambda_4 c_4+\lambda_5 c_5)\ \frac{1}{2}\Bigg[ I_s(1) + \frac{1}{3}I_s\left(\dfrac{m_I^2}{m_R^2}\right) +  \frac{2}{3}I_s\left(\dfrac{m_{S^\pm}^2}{m_R^2}\right)\Bigg]\Bigg\}\\ 
F_I^{gg} &= \Bigg\{
-\eta_U s_U\ I_q\left(\dfrac{m_t^2}{m_I^2}\right)
+\eta_Ds_D\ I_q\left(\dfrac{m_b^2}{m_I^2}\right) \nonumber \\
&-\dfrac{9}{4}\dfrac{v^2}{m_I^2}\ (\lambda_4 s_4+\lambda_5 s_5)\ \dfrac{1}{2}\Bigg[ I_s(1)+ \frac{1}{3}I_s\left(\dfrac{m_R^2}{m_I^2}\right) +  \frac{2}{3}I_s \left(\dfrac{m_{S^\pm}^2}{m_I^2}\right)\Bigg]\Bigg\} \\ \nonumber\\
\tilde F_R^{gg} &= \left[
-\eta_U s_U\ \dfrac{m_t^2}{m_R^2}\  f\left(\dfrac{m_t^2}{m_R^2}\right)
-\eta_D s_D\ \dfrac{m_b^2}{m_R^2}\ f\left(\dfrac{m_b^2}{m_R^2}\right)\right]\\ \nonumber\\
\tilde F_I^{gg} &=\left[
-\eta_U c_U\ \dfrac{m_t^2}{m_I^2}\ f\left(\dfrac{m_t^2}{m_I^2}\right)
+\eta_D c_D\ \dfrac{m_b^2}{m_I^2}\ f\left(\dfrac{m_b^2}{m_I^2}\right)\right]
\end{align}
where $I_q$, $I_s$ and $f$ are familiar from $Hgg$ effective couplings and are given below.

Imposing custodial symmetry, the scalar loops only contribute to an $S_R gg$ coupling through $F^{gg}_R$. Imposing CP symmetry $S_R$ ($S_I$) are pure scalar (pseudo-scalar) and therefore only the factors $F_R^{gg}$ and $\tilde F^{gg}_I$ are not-zero. The bottom-quark loops are much suppressed with respect to the top-quark loops unless $\eta_D>>\eta_U$. In the limit of CP conservation and $m_b=0$, these results agree with \cite{Gresham:2007ri} except for the sign in front of the factor $\dfrac{9}{4}$. This sign, however, is of no consequence for phenomenology as $\lambda_{4,5}$ can have either sign. \footnote{When CP violation is included we find the following errors in \cite{He:2011ws}: the factors $\tilde F_{R,I}^{gg}$ are a factor of two too large in \cite{He:2011ws}; the function $I_s(z)$ in Eq.~2.6 of \cite{He:2011ws} contains an incorrect overall factor of $z$ which is inconsequential in the limit of degenerate scalars.There is also a typo in Eq.~6 of \cite{Hayreter:2017wra}, where there should be a minus sign in the term with $\eta_U$ in $F_I$, corresponding to $\tilde F_I^{gg}$ here.}

\subsection{\boldmath\underline{$S_{R,I}\to \gamma\ g$}}

\begin{figure}[ht]
\centering
\includegraphics[scale=0.45]{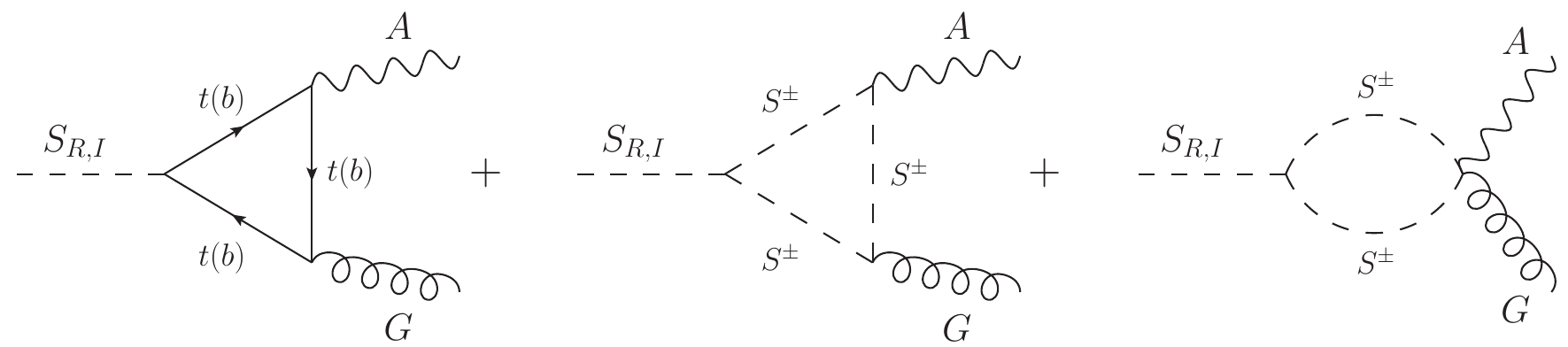}
\caption{One-loop diagrams contributing to the factors appearing in Eq.~(\ref{defff}) for the  $S_{R,I}\gamma g$ coupling.}
\label{f:sgammag}
\end{figure}

The diagrams leading to $S\gamma g$ effective vertices are shown in Figure~\ref{f:sgammag} and the resulting form factors are given by
\begin{align}
F_R^{\gamma g} &= \left[
\eta_U c_U\ I_q\left(\dfrac{m_t^2}{m_R^2}\right)
-\eta_Dc_D\ \dfrac{1}{2}I_q\left(\dfrac{m_b^2}{m_R^2}\right)
 -\dfrac{9}{4}\dfrac{v^2}{m_R^2}\ (\lambda_4 c_4-\lambda_5 c_5)\ \frac{1}{2}I_s\left(\dfrac{m_{S^\pm}^2}{m_R^2}\right)\right]\\ \nonumber\\
F_I^{\gamma g} &= \left[
-\eta_U s_U\ I_q\left(\dfrac{m_t^2}{m_I^2}\right)
-\eta_Ds_D\ \dfrac{1}{2}I_q\left(\dfrac{m_b^2}{m_I^2}\right)
 +\dfrac{9}{4}\dfrac{v^2}{m_I^2}\ (\lambda_4 s_4-\lambda_5 s_5)\  \frac{1}{2} I_s\left(\dfrac{m_{S^\pm}^2}{m_I^2}\right)\right]\\ \nonumber\\
\tilde F_R^{\gamma g} &=\left[
-\eta_U s_U\ \dfrac{m_t^2}{m_R^2}\ f\left(\dfrac{m_t^2}{m_R^2}\right)
+\eta_D s_D\ \dfrac{1}{2}\ \dfrac{m_b^2}{m_R^2}\ f\left(\dfrac{m_b^2}{m_R^2}\right)\right]\\ \nonumber\\
\tilde F_I^{\gamma g} &= \left[
-\eta_U c_U\ \dfrac{m_t^2}{m_I^2}\ f\left(\dfrac{m_t^2}{m_I^2}\right)
-\eta_D c_D\ \dfrac{1}{2}\ \dfrac{m_b^2}{m_I^2}\ f\left(\dfrac{m_b^2}{m_I^2}\right)\right] 
\end{align}
We note that the scalar loop contribution to $S_R\to \gamma g$ vanishes in the custodial symmetry limit. In the custodial and CP symmetry limits, the scalar loops do not affect these processes. 

\subsection{\boldmath\underline{$S_{R,I}\to Z\ g$}}
\begin{figure}[h]
\centering
\includegraphics[scale=0.45]{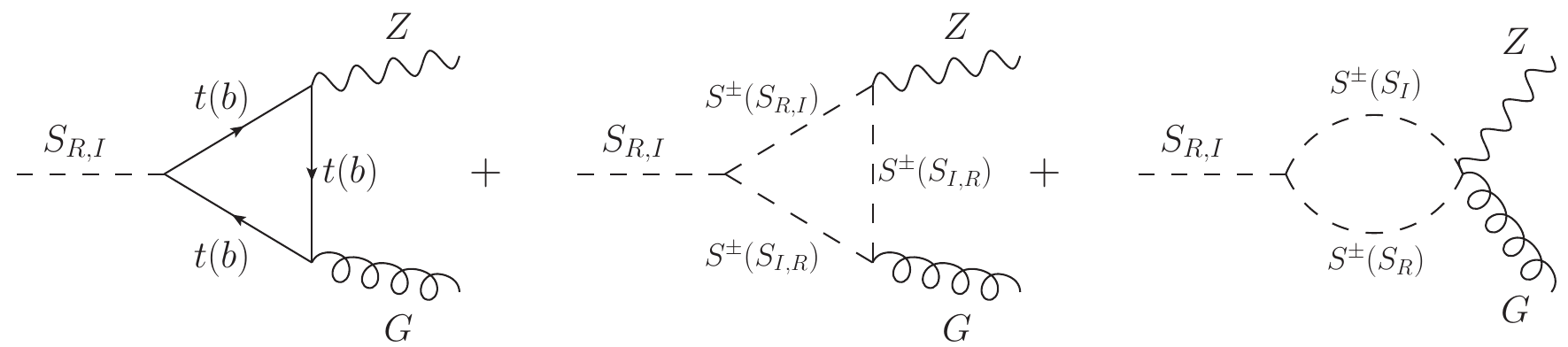}	
\caption{One-loop diagrams contributing to the factors appearing in Eq.~(\ref{defff}) for the $S_{R,I}Zg$ coupling.}
\label{f:szg}
\end{figure}

In this case, the relevant diagrams are shown in Figure~\ref{f:szg} and result in form factors given by

\begin{align}\hspace*{-1.0cm}
F_R^{Zg} = \Bigg\{
&-\eta_U c_U\ \dfrac{(3c_W^2-5s_W^2)}{s_{2W}} \Bigg[I_1\left(\dfrac{m_t^2}{m_R^2},\dfrac{m_t^2}{m_Z^2}\right)-I_2\left(\dfrac{m_t^2}{m_R^2},\dfrac{m_t^2}{m_Z^2}\right)\Bigg] \nonumber \\
&+\eta_D c_D\ \frac{(3c_W^2-s_W^2)}{s_{2W}} \Bigg[I_1\left(\dfrac{m_b^2}{m_R^2},\dfrac{m_b^2}{m_Z^2}\right)-I_2\left(\dfrac{m_b^2}{m_R^2},\dfrac{m_b^2}{m_Z^2}\right)\Bigg] \nonumber \\
&-\frac{9}{4}\frac{v^2}{m_{S^\pm}^2}\frac{1}{ t_{2W}}(\lambda_4c_4-\lambda_5c_5)\Bigg[
I_1\left(\dfrac{m_{S^\pm}^2}{m_R^2},\dfrac{m_{S^\pm}^2}{m_Z^2}\right)\Bigg]
\Bigg\} 
\end{align}
\begin{align}\hspace*{-1.0cm}
F_I^{Zg} = \Bigg\{
&\eta_U s_U\ \dfrac{(3c_W^2-5s_W^2)}{s_{2W}} \Bigg[I_1\left(\dfrac{m_t^2}{m_I^2},\dfrac{m_t^2}{m_Z^2}\right)-I_2\left(\dfrac{m_t^2}{m_I^2},\dfrac{m_t^2}{m_Z^2}\right)\Bigg] \nonumber \\
&+\eta_D s_D\ \frac{(3c_W^2-s_W^2)}{s_{2W}} \Bigg[I_1\left(\dfrac{m_b^2}{m_I^2},\dfrac{m_b^2}{m_Z^2}\right)-I_2\left(\dfrac{m_b^2}{m_I^2},\dfrac{m_b^2}{m_Z^2}\right)\Bigg] \nonumber \\
&+\frac{9}{4}\frac{v^2}{m_{S^\pm}^2}\frac{1}{ t_{2W}}(\lambda_4s_4-\lambda_5s_5)\Bigg[
I_1\left(\dfrac{m_{S^\pm}^2}{m_I^2},\dfrac{m_{S^\pm}^2}{m_Z^2}\right)\Bigg]
\Bigg\} 
\end{align}
\begin{align}\hspace*{-1.0cm}
\tilde F_R^{Zg}  =  \frac{1}{(m_R^2-m_Z^2)}\Bigg\{ -\eta_U s_U\ \dfrac{(3c_W^2-5s_W^2)}{s_{2W}}\  m_t^2 \Bigg[ f\left(\dfrac{m_t^2}{m_R^2}\right)-f\left(\dfrac{m_t^2}{m_Z^2}\right)\Bigg] \nonumber \\ \nonumber \\ 
+\eta_D s_D\ \dfrac{(3c_W^2-s_W^2)}{s_{2W}}\  m_b^2 \Bigg[ f\left(\dfrac{m_b^2}{m_R^2}\right)-f\left(\dfrac{m_b^2}{m_Z^2}\right)\Bigg]\Bigg\}
\end{align}

\begin{align}\hspace*{-1.0cm}
\tilde F_I^{Zg}  =   \frac{1}{(m_I^2-m_Z^2)}\Bigg\{ -\eta_U c_U\ \dfrac{(3c_W^2-5s_W^2)}{s_{2W}}\  m_t^2 \Bigg[ f\left(\dfrac{m_t^2}{m_I^2}\right)-f\left(\dfrac{m_t^2}{m_Z^2}\right)\Bigg] \nonumber \\ \nonumber \\ 
-\eta_D c_D\ \dfrac{(3c_W^2-s_W^2)}{s_{2W}}\  m_b^2 \Bigg[ f\left(\dfrac{m_b^2}{m_I^2}\right)-f\left(\dfrac{m_b^2}{m_Z^2}\right)\Bigg]\Bigg\}
\end{align}
The functions $I_{1,2}(x,y)$ already appear in the scalar contributions to $H\to Z\gamma$ \cite{Chen:2013vi} and are given below. 
Once again, the scalar loop contributions to $S_R\to Zg$ vanish if custodial symmetry is imposed and those to to $S_I\to Zg$ vanish when CP symmetry is imposed.

\subsection{\boldmath\underline{$S^+\to W^+\ g$}}
\begin{figure}[h]
\centering
\includegraphics[scale=0.45]{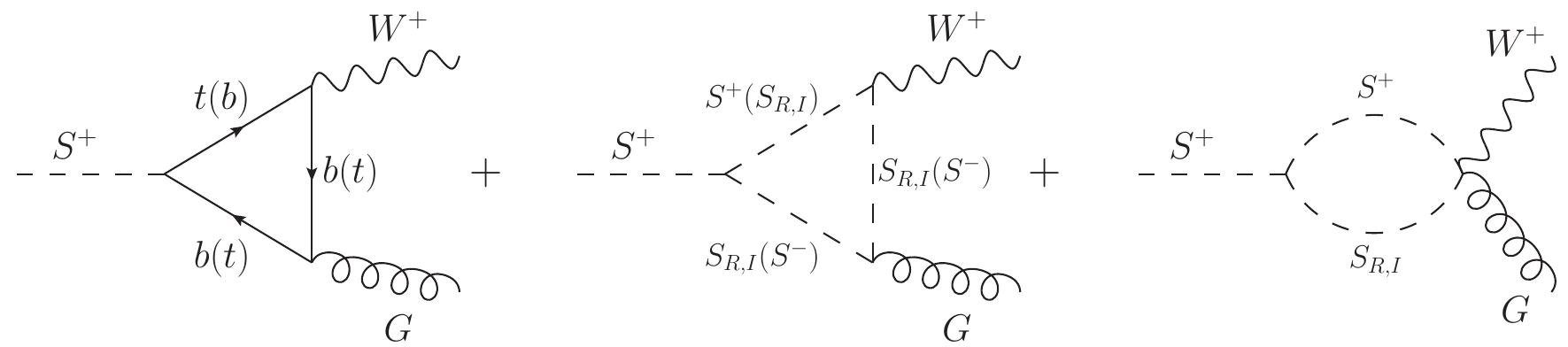}
\caption{One-loop diagrams contributing to the factors appearing in Eq.~(\ref{defff}) for the  $S^\pm W^\mp g$ coupling.}
\label{f:swg}
\end{figure}	

Finally, the diagrams for $S^+\to W^+\ g$  are shown in Figure~\ref{f:swg}. The complete result is rather cumbersome and quite complicated therefore to simplify the calculation we preferred to choose the case where $m_{S^\pm}=m_I$ and we present it in the appendix. It simplifies 
considerably if $m_b\to 0$ and we treat the scalars as degenerate. In this case, and separating the quark  and  scalar loop contributions into $F^{Wg}=F_q^{Wg}+F_S^{Wg}$ we find,

\begin{itemize}
\item the quark loop contributions in the limit $m_b\to 0$ become
\begin{eqnarray}
F_q^{Wg} &=& \frac{3\eta_Ue^{i\alpha_U}|V_{tb}|^2m_t^2}{2m_I^2(m_I^2-m_W^2)^2 s_W}\left[
\frac{1}{2}m_I^2(m_I^2-2m_t^2-m_W^2)\cdot \right.
 \nonumber \\
&&\left.
\left(2{\rm Li}_2\left(\frac{m_t^2}{m_I^2}\right)-2{\rm Li}_2\left(\frac{m_t^2}{m_W^2}\right)
+\log\left(\frac{m_W^2}{m_I^2}\right)\log\left(\frac{m_t^4}{m_I^2m_W^2}\right)
\right)
-2m_I^2(m_I^2-m_W^2) \right.\nonumber\\
&+&\left.2m_I^2(m_t^2-m_W^2)\log\left(\frac{m_t^2-m_W^2}{m_t^2}\right)
+2m_W^2(m_I^2-m_t^2)\log\left(\frac{m_t^2-m_I^2}{m_t^2}\right)\right]
\end{eqnarray}
\item  the scalar loop contributions when all scalars are degenerate and $m_W<<m_S^\pm$ become
\begin{eqnarray}
F_S^{Wg} &=&\frac{9v^2}{4m_S^2\  s_W}I_s(1)(\lambda_4-\lambda_5)\left(1+{\cal O}\left(\frac{m_W^2}{m_S^2}\right)\right)
\end{eqnarray}
In the same limit,
\begin{eqnarray}\hspace*{-1.0cm}
\tilde F^{Wg} = \frac{1}{s_W}\Bigg\{
-\frac{3i \eta_U|V_{tb}|^2m_t^2 }{4(m_I^2-m_W^2)}\left(2{\rm Li}_2\left(\frac{m_t^2}{m_I^2}\right)-2{\rm Li}_2\left(\frac{m_t^2}{m_W^2}\right)
+\log\left(\frac{m_W^2}{m_I^2}\right)\log\left(\frac{m_t^4}{m_I^2m_W^2}\right)
\right)\Bigg\}
\end{eqnarray}
\end{itemize}

\subsection{Loop functions}
The functions appearing in the above results are given by
\begin{eqnarray}
I_q(x) &= &2x + x(4x-1)f(x),\nonumber\\
 I_s(x) &=& -(1+2x f(x)),\nonumber\\
I_1(x,y)&=&\dfrac{2xy}{(x-y)}-\dfrac{4x^2y^2}{(x-y)^2}[f(x)-f(y)]+\dfrac{4x^2y}{(x-y)^2}[g(x)-g(y)] \nonumber \\
I_2(x,y)&=&\dfrac{xy}{(x-y)}[f(x)-f(y)] 
\end{eqnarray}
where
\begin{eqnarray}
f(x)=\left\lbrace
\begin{array}{ccc}
\dfrac{1}{2}\left(\ln\left(\dfrac{1+\sqrt{1-4x}}{1-\sqrt{1-4x}}\right)-i\pi\right)^2 & & \text{for }x<\dfrac{1}{4}\\ & & \\
-2\left(\arcsin\left(\dfrac{1}{2\sqrt{x}}\right)\right)^2 & & \text{for }x>\dfrac{1}{4}
\end{array}\right.
\end{eqnarray}
and
\begin{eqnarray}
g(x)=\left\lbrace
\begin{array}{ccc}
\dfrac{1}{2}\sqrt{1-4x}\left(\ln\left(\dfrac{1+\sqrt{1-4x}}{1-\sqrt{1-4x}}\right)-i\pi\right) & & \text{for }x<\dfrac{1}{4}\\ & & \\
\sqrt{4x-1}\arcsin\left(\dfrac{1}{2\sqrt{x}}\right) & & \text{for }x>\dfrac{1}{4}
\end{array}\right.
\label{mygx}
\end{eqnarray}
The special value appearing for degenerate scalar masses, $I_s(1)=\dfrac{\pi^2}{9}-1$.

\section{Decay widths in different scenarios}

The decay widths are given in terms of the couplings defined in Eq.~\ref{defff} by 
\begin{eqnarray}
\Gamma(S_{R,I}\to gg) &=& \frac{5}{12\pi}\left(\frac{\alpha_s}{8\pi v}\right)^2 m_{R,I}^3\left(|F_{R,I}^{gg}|^2+|\tilde F_{R,I}^{gg}|^2\right)\nonumber \\ \nonumber \\
\Gamma(S_{R,I}\to \gamma g) &=& \frac{1}{8\pi}\left(\frac{\alpha \alpha_s}{(3\pi v)^2}\right)m_{R,I}^3\left(|F_{R,I}^{\gamma g}|^2+|\tilde F_{R,I}^{\gamma g}|^2\right)\nonumber \\ \nonumber \\
\Gamma(S_{R,I}\to Z g) &=& \frac{1}{8\pi}\left(\frac{\alpha \alpha_s}{(12\pi v)^2}\right)\frac{(m_{R,I}^2 -m_Z^2)^3}{m_{R,I}^3}\left(|F_{R,I}^{Zg}|^2+|\tilde F_{R,I}^{Zg}|^2\right)\nonumber \\ \nonumber \\
\Gamma(S^\pm\to W^\pm g) &=& \frac{1}{8\pi}\left(\frac{\alpha \alpha_s}{(12\pi v)^2}\right)\frac{(m_{S^+}^2-m_W^2)^3}{m_{S^+}^3}\left(|F_{R,I}^{Wg}|^2+|\tilde F_{R,I}^{Wg}|^2\right)
\end{eqnarray}
Numerically, these result in small branching ratios that are negligible for phenomenology except in special cases.  

\subsection{General remarks on parameter space}

There are two cases in which the $Vg$ modes are important and they both rely on mechanisms to suppress scalar decay into heavy quarks. 
\begin{itemize}
\item The $S_{R,I}$ neutral resonances will decay predominantly into top pairs and $S^\pm$ will decay predominantly into top-bottom pairs if those channels are  kinematically available. This means that the loop induced modes become important for the mass ranges
\begin{eqnarray}
100{\rm~GeV} \lesssim m_{R,I} \lesssim 350{\rm~GeV}, && 100{\rm~GeV} \lesssim m_{S^\pm} \lesssim 175{\rm~GeV}.
\end{eqnarray}
The lower limit corresponds approximately to the LEP exclusion for scalar pair production \cite{Burgess:2009wm}. Scalar decay into two jets through couplings to the lighter quarks are not suppressed in these ranges and in some instances will dominate as shown below.

\item The decays to $t\bar{t}$ or $t\bar{b}$ can also be suppressed with very small values of $\eta_U$. This would also suppress the production mechanism for a single scalar, but would not affect the cross-section for pair-production which depends only on the QCD coupling constant and at 13 TeV is approximately 0.2~pb. \cite{Gresham:2007ri,Hayreter:2017wra}. Reducing $\eta_U$ also reduces the top-quark contribution to the loop decays, which is dominant in most cases.

\item When  $\eta_U$ is small,  $S_{R,I}$ will also decay predominantly into bottom pairs unless $\eta_D$ is also small.  It is possible for the color octet scalars to have very small Yukawa couplings, or to introduce discrete symmetries  \cite{Pois:1993ay} or additional scalar multiplets \cite{Cheng:2016tlc} so that they vanish, resulting in  fermiophobic scenarios \cite{Miralles:2019uzg}.

\item Large values of $\lambda_{4,5}$,  that obey the condition $\lambda_4=\lambda_5^\star$, can significantly affect the $gg$ channel but not the other ones. 

\item It is possible to enhance the $\gamma g$ or $Zg$ modes relative to the $gg$ mode if the sign of $\lambda_4$ results in destructive interference with the fermion loops for $gg$.

\item Kinematic windows can also be used to suppress decays between the different scalars. For example, choosing  $\lambda_2=0$ results in degeneracy between $S_I$ and $S_R$ thus preventing (on-shell) decays between them.

\end{itemize}

\subsection{Scalar potential with $\lambda_5\neq \lambda_4^\star$}

One of the conditions arising from imposing custodial symmetry is 
$\lambda_4=\lambda_5^\star$. These parameters, however, do not affect the $W,Z$ masses until at least the two-loop level so constraints from the $\rho$ parameter are much weaker than those on $\lambda_{2,3}$. Entertaining this possibility permits a suppression of multijet modes (via the the $S \to gg$ decay) in favor of the $\gamma g$ and $Zg$ modes. We illustrate this for two cases:
\begin{itemize}
\item $CP$ conserving scenario with $\lambda_5=-\lambda_4$. This completely removes the scalar loop contributions in the $S_{R,I}\to gg$ modes while enhancing them in $S_{R,I}\to \gamma g,Z g$ as well as in $S^\pm \to W^\pm g$.
\item $CP$ violating scenario with $|\lambda_5|=|\lambda_4|$ but both have a phase of $\pi/2$. This  also removes the scalar loop contributions in the $S_{R,I}\to gg$  and enhances them in $S_{R,I}\to \gamma g,Z g$, this time via $CP$ violating contributions.
\end{itemize}

\subsection{Parameter choices}

For our numerical results we will keep $\eta_U\leq 5$, below its unitarity constraint \cite{He:2013tla}. We will choose a smaller $\eta_D$, typically $\eta_D\leq 1$ as we do not want to enhance the $b\bar{b}$ decay modes. To illustrate the fermiophobic scenario we set $\eta_U=0$ but keep $\eta_D$ small to show the interplay between the different modes. In all cases 
we keep $|\lambda_{4,5}|\leq 10$, which is below its tree level unitarity constraint \cite{He:2013tla}  and within the range of next to leading order unitarity constraints \cite{Cheng:2018mkc}. We select a few benchmarks for non-zero CP phases in the scalar potential.

\section{Numerical Results}

\subsection{Low scalar mass window}

\begin{figure}[h]
\centering\includegraphics[scale=0.5]{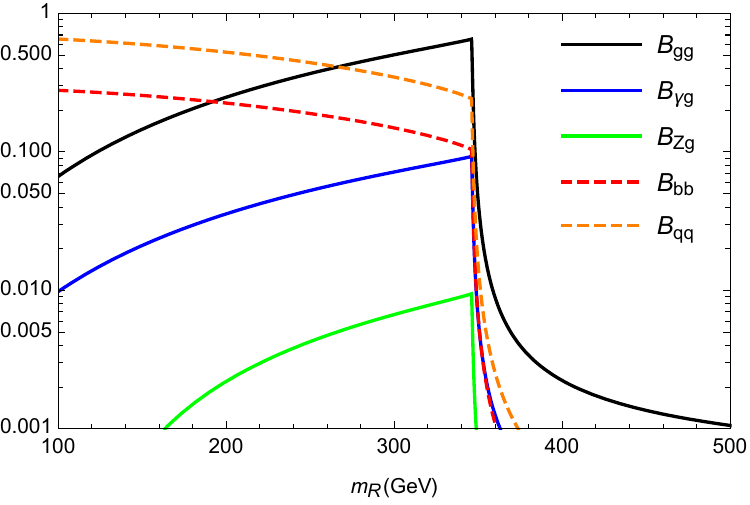}\hspace{.5in}\includegraphics[scale=0.5]
{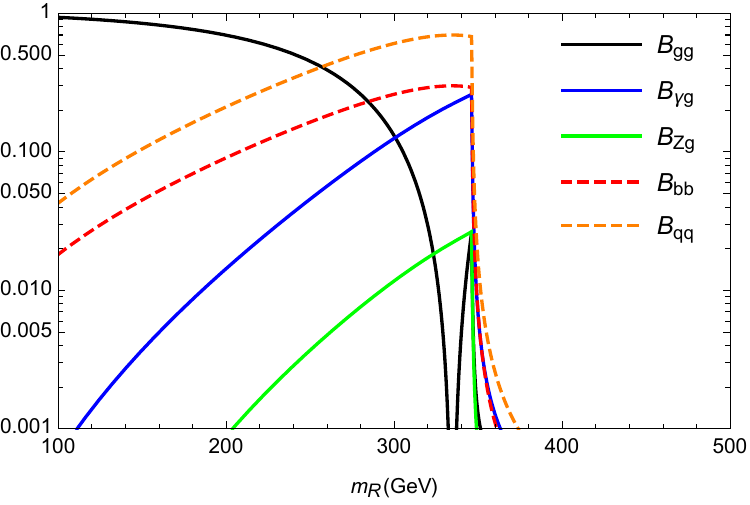}
\caption{Branching ratios for $S_R$ decay in the fermio-philic scenario $\eta_U=5$, $\eta_D=1$  (left panel: $\lambda_{4,5}=0$, right panel: $\lambda_{4}=\lambda_{5}=-10$.  Above the $t\bar{t}$ threshold,  ${\cal B}(S_R\to t\bar{t})$ rises rapidly to 100\%.}
\label{f:brsr}
\end{figure}

\begin{figure}[h]
\centering\includegraphics[scale=0.5]{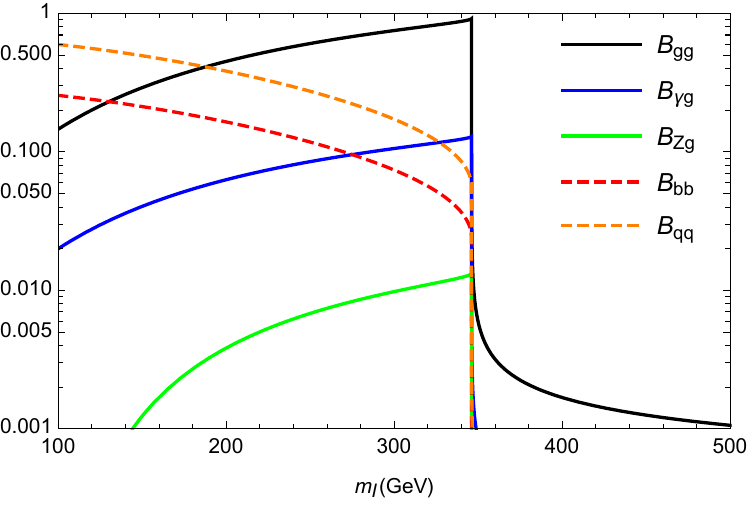}\hspace{.5in}\includegraphics[scale=0.5]
{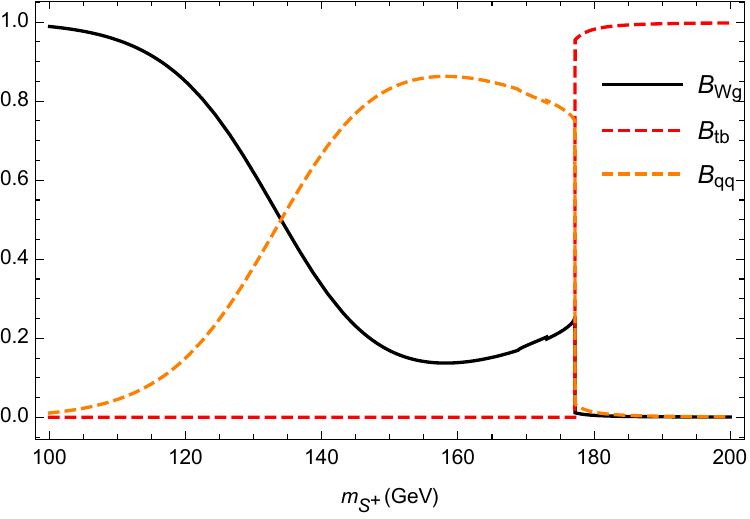}
\caption{Branching ratios for $S_I$ left panel ($S^\pm$ right panel) decay for parameter values $\eta_U=5$ and $\eta_D=1$. Above the $t\bar{t}$ threshold,  ${\cal B}(S_I\to t\bar{t})$ rises rapidly to 100\%.}
\label{f:brsp}
\end{figure}

When the Yukawa couplings are of order one, the decay modes into third generation fermions are completely dominant. In this case the loop induced $Vg$ modes only become important for sufficiently low scalar masses. 
To illustrate this scenario we select $\eta_U=5$, $\eta_D=1$ in Figures~\ref{f:brsr}~and~\ref{f:brsp} where the $Vg$ loop modes are shown as solid lines and tree-level $q\bar{q}$ modes as dashed lines. For masses $m_R$ above $2m_t$ the ${\cal B}(S_R\to t\bar{t})\approx 1$ unless $\eta_U \lesssim \eta_D m_b/m_t$, in which case ${\cal B}(S_R\to b\bar{b})$ becomes comparable. The choice $\eta_U=5$ for $m_R$ below $2m_t$ enhances the rate into $Vg$ modes relative to tree-level $b\bar{b}$ modes through the top-quark loop contribution.

The relevant branching ratios for $S_R$ decay are shown in Figure~\ref{f:brsr}. The figure illustrates the interplay between $gg$, $qq$ (light-quarks) and $b\bar{b}$ modes which are the most important ones in this case.  On the left panel we set $\lambda_{4,5}=0$, so the contributions to $Vg$ modes are only from quark loops. For the right panel we used  $\lambda_{4}=\lambda_{5}=-10$ in order to enhance the contribution of the scalar loops, and with the sign chosen to suppress the $gg$ mode near the $t\bar{t}$ threshold. In this case the $\gamma g$ mode can reach branching ratios near 12\% and the $Zg$  mode near 1\%.

The corresponding decay modes for $S_I$ are shown in the left panel of Figure~\ref{f:brsp}. As mentioned before,  $\lambda_{4,5}$ do not affect $S_I\to gg$ if custodial symmetry is imposed ($\lambda_5=\lambda_4^\star$), and they do not affect $S_I\to (\gamma/Z) g$ if CP symmetry is imposed. The branching ratios shown in Figure~\ref{f:brsp} will thus vary only with the ratio $\eta_U/\eta_D$. The right panel of the same figure illustrates decay modes of the charged scalar, for which  $Wg$ mode can dominate for values of $m_{S^\pm}$ below the $t\bar{b}$ threshold.
The $qq$ light-quark mode below $t\bar{b}$ threshold is dominated by $c\bar{b}$ quarks. 

\subsection{Fermio-phobic scenario}

\begin{figure}[h]
\centering\includegraphics[scale=0.5]{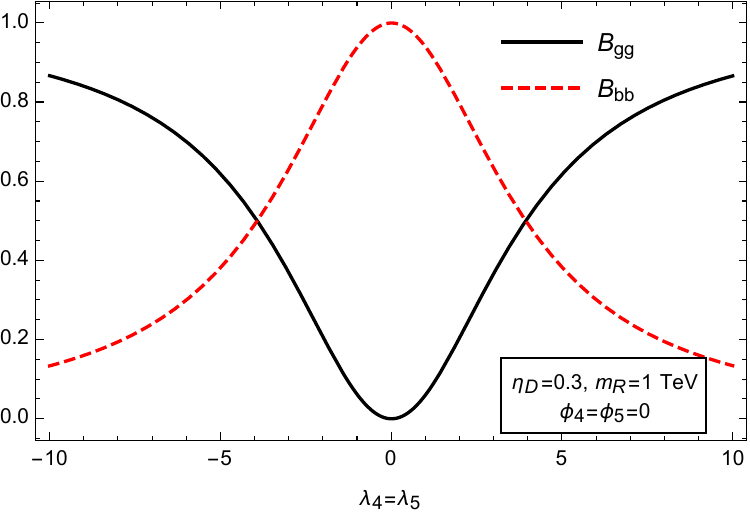}\hspace{.5in}\includegraphics[scale=0.5]
{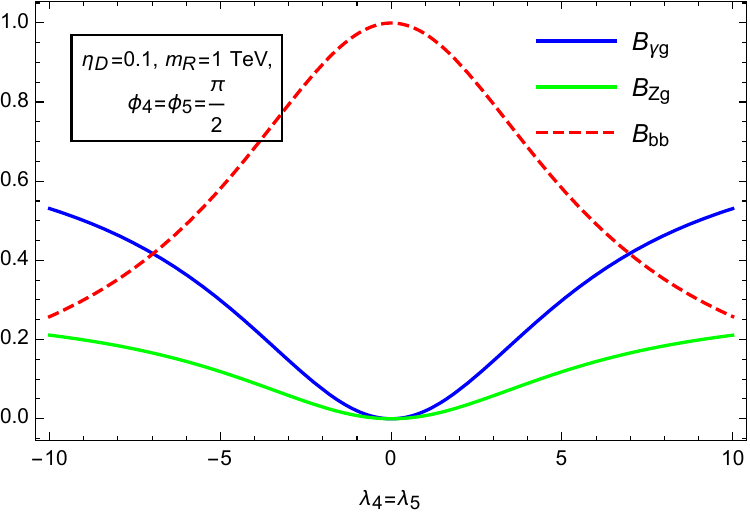}
\centering\includegraphics[scale=0.5]{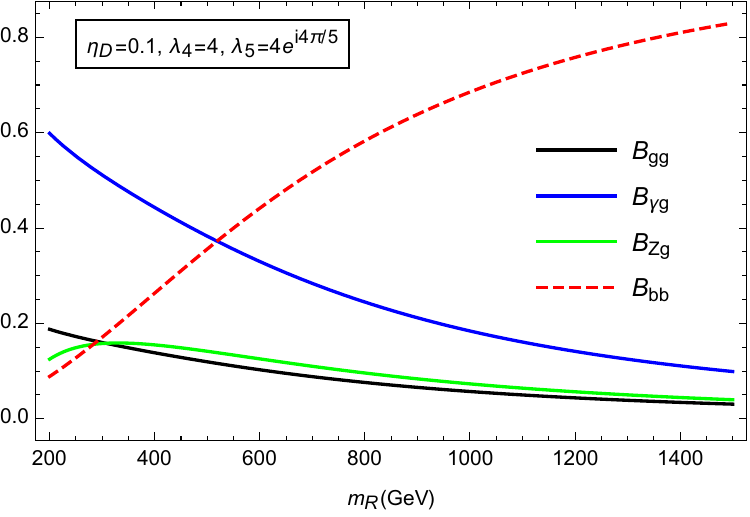}\hspace{.5in}\includegraphics[scale=0.5]
{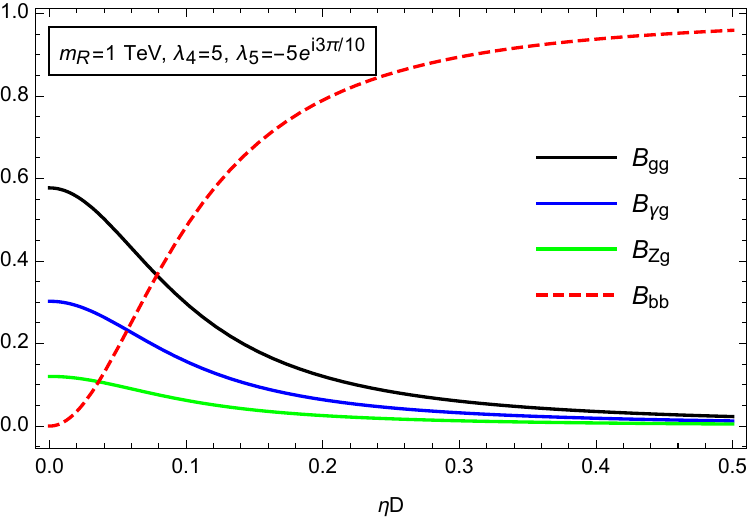}
\centering\includegraphics[scale=0.5]{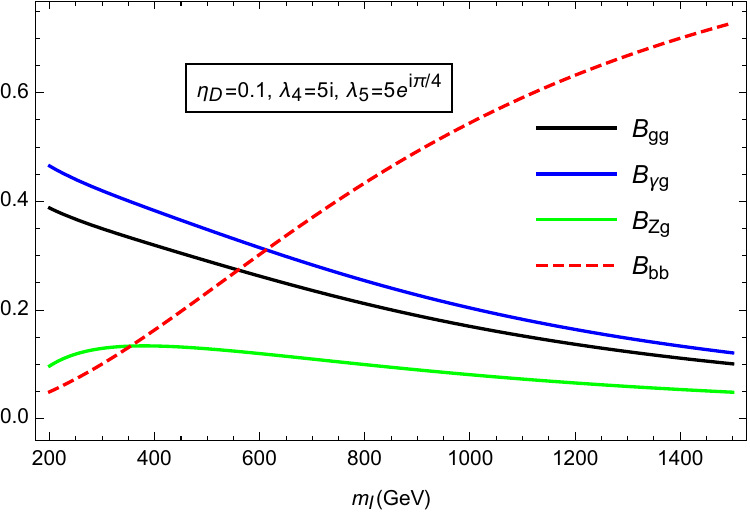}\hspace{.5in}\includegraphics[scale=0.5]{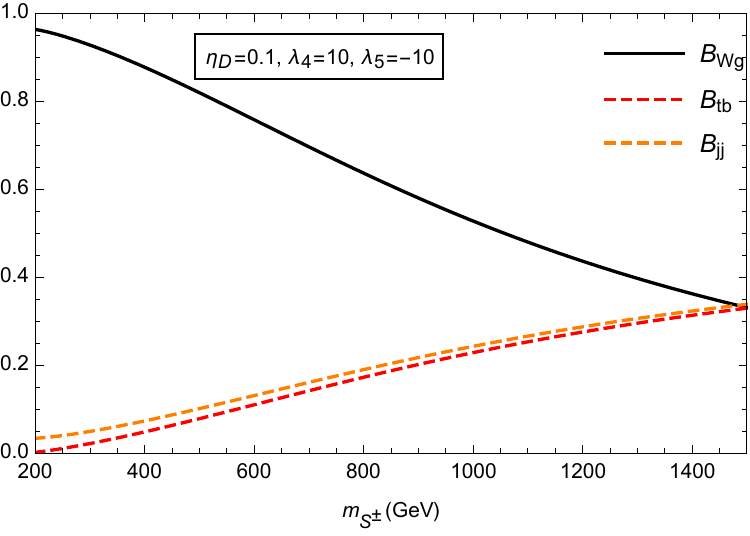}
\caption{Illustrative parameter points for $S_R$ decay modes, top four panels; $S_I$ decay modes, bottom left; and $S^\pm$ decay modes, bottom right. In all cases $\eta_U=0$. }
\label{f:srmix}
\end{figure}

In the limit of vanishing Yukawa couplings, $\eta_{U,D}=0$,  the tree-level decays of $S$ into fermion pairs vanish, and the one-loop modes are driven by scalar loop contributions. For the charged scalar we have in this case  ${\cal B}(S^\pm \to W^\pm g)\approx1$ provided $\lambda_5\neq\lambda_4$. The situation for the 
neutral scalars is more complicated and we show some limiting cases in Table~\ref{t:ff}.
\begin{table}[htp]
\begin{center}
\begin{tabular}{|c|c|c|c|c|c|}
\hline
$\phi_4=\phi_5=0$ & $\lambda_5=\lambda_4$ & $\lambda_5=-\lambda_4$ &
$\phi_4=\phi_5=\frac{\pi}{2}$ & $\lambda_5=\lambda_4$ & $\lambda_5=-\lambda_4$ \\ \hline
${\cal B}(S_R\to gg)$& 1& 0  &
${\cal B}(S_I\to gg)$& 1& 0 \\
${\cal B}(S_R\to \gamma g)$& 0& 71.5\% &
${\cal B}(S_I\to \gamma g)$& 0& 71.5\% \\
${\cal B}(S_R\to Z g)$& 0& 28.5\% &
${\cal B}(S_I\to Z g)$& 0& 28.5\% \\ \hline
\end{tabular}
\end{center}
\caption{Sample parameter points with $\eta_{U,D}=0$.}
\label{t:ff}
\end{table}%
For small, but non-zero $\eta_D$, decays into $b$ quarks compete with the loop-induced modes. We illustrate this interplay for selected parameter points in Figure~\ref{f:srmix}, where $\eta_U=0$ in all cases. 
The top two panels show $S_R$ branching ratios: on the left we take $\lambda_4=\lambda_5$ with no phases, which removes the $\gamma g$ and $Zg$ modes. The ratio $\Gamma(S_R\to gg)/\Gamma(S_R\to bb)$ increases with increasing $\lambda_4/\eta_D$; on the right we still take $\lambda_4=\lambda_5$ but allow a $90^\circ$ phase which removes the $gg$ mode in favor of $\gamma g$ and $Zg$. The two panels in the center  row illustrate the dependence on $m_R$ and $\eta_D$ for certain fixed choices of $\lambda_{4,5}$, again for  $S_R$ decay. Finally in the bottom row, we show branching ratios for $S_I$ decay on the left panel and for $S^\pm$ decay on the right panel. In both cases we illustrate the dependence on the scalar mass for fixed choices of $\eta_D$ and $\lambda_{4,5}$.
The message from Figure~\ref{f:srmix} is that the $Vg$ modes can be dominant in fermiophobic scenarios, and that their relative importance varies across the parameter space. The study of $Vg$ modes is therefore necessary to fully constrain models with new colored scalars.

\subsection{Comparison with the literature}

A recent analysis of $gg$, $\gamma g$ and $Z g$ modes at LHC appeared in  \cite{Cacciapaglia:2020vyf}. This study is quite different from ours, as it concerns models where the effective vertices arise through a Wess-Zumino term in composite models. They offer a parameterization of the effective vertices that we can use to compare to our results, they write
\begin{eqnarray}
{\cal L}_\Phi \supset i C_t \frac{m_t}{f_\Phi}\Phi^a \bar{t}\gamma_5 
\frac{\lambda^a}{2}t+\frac{\alpha_s\kappa_g}{8\pi f_\Phi}\Phi^a\epsilon^{\mu\nu\rho\sigma}\left[
\frac{1}{2}d^{abc}G^b_{\mu\nu}G^c_{\rho\sigma}+\frac{e\kappa_\gamma}{g_s\kappa_g}G^a_{\mu\nu}F_{\rho\sigma}-\frac{e\tan_W \kappa_Z}{g_s\kappa_g}G^a_{\mu\nu}Z_{\rho\sigma}\right]
\label{lcomp}
\end{eqnarray}
where $\Phi$ is assumed to be a composite octet pseudoscalar with mass scale $f_\Phi$. Our results are given in the form of Eq.~\ref{defff}, indicating that the MW model at one-loop produces the effective couplings in Eq.~\ref{lcomp} as
\begin{eqnarray}
C_t=\frac{f_\Phi}{v}\eta_U,~~\kappa_g =\frac{f_\Phi}{v} \tilde{F}_I^{gg},~~
\kappa_\gamma =\frac{8}{3} \frac{f_\Phi}{v} \tilde{F}_I^{\gamma g} ,~~
\kappa_Z = \frac{2}{3}\cot_W \frac{f_\Phi}{v} \tilde{F}_I^{Z g} 
\end{eqnarray}
if we identify the neutral pseudoscalar $S_I$ with the composite $\Phi$. We find, in agreement with \cite{Cacciapaglia:2020vyf}, that the loop-induced modes are phenomenologically relevant when $C_t<<\kappa_g$, which in the MW model corresponds to a dominance of the scalar loops due to vanishing (or very small) $\eta_U$. 

In addition, the results of \cite{Cacciapaglia:2020vyf} are presented in terms of the ratio $\kappa_\gamma/\kappa_g$. In the MW model, this ratio can be tuned from 0 to 1 as can be seen in Table~\ref{t:ff}. In the regime of interest for the $\gamma g$ and $Zg$ modes, single production of $S$ at the LHC is unobservably small so any bounds would come from QCD pair production of $SS$ followed by decays into $\gamma/Z j$.   The case of $SS$ pair production followed by multi top/bottom/jet decays for the MW model was already studied by us in \cite{Hayreter:2017wra}. In that paper we found significant constraints already exist on the $SS$ pair production followed by decays into dijet pairs. Those results suggest that the key issue in constraining the $Vj$ modes will be the ability of ATLAS and CMS to reconstruct these final states. The theoretical study in \cite{Cacciapaglia:2020vyf}  estimates the relevant backgrounds and finds regions in the $(m_\Phi,~{\cal B}(\Phi\to g\gamma))$ plane that can be covered by $jjj\gamma$ and $jj\gamma\gamma$ searches at 14~TeV. Based on those results we can conclude that these modes can also be used to constrain the MW model.

\section{Summary and conclusion}

We have presented for the first time explicit one-loop results for the decay modes of a scalar color octet $S^0\to \gamma g$, $S^0\to Zg$ and $S^\pm \to W^\pm g$ in the MW model. These results arise from heavy quark loops proportional to the Yukawa couplings of the colored scalars and from colored scalar loops proportional to the triple scalar coupling in the potential. We have further identified the regions of parameter space where these modes become important. These regions correspond to fermiophobic scenarios and/or to low scalar masses for which decays into heavy quarks are kinematically forbidden.

\section*{Acknowledgments} We thank Thomas Flacke for pointing out these vertices did not appear in the literature. We also thank Vladyslav Shtabovenko for very useful discussions about the automation of loop calculations with the FeynCalc package.

\appendix

\section{Complete loop factors for $S^\pm\to W^\pm g$}

The complete expressions we obtain in this case can be written in terms of the Passarino-Veltman function $C_0$ as follows,
\begin{eqnarray}
F^{Wg} &=& \frac{1}{(m_I^2-m_W^2)^2\ s_W}\Bigg\{
3|V_{tb}|^2\Bigg[(m_t^2\ \eta_U e^{i\alpha_U}-m_b^2\ \eta_D e^{-i\alpha_D})\nonumber \\ \nonumber \\
&\times&\Bigg((m_W^2-m_b^2-m_t^2)\ I^+\left(\frac{m_b^2\ m_t^2}{(m_W^2-m_b^2-m_t^2)^2}\right)-\dfrac{m_W^2}{m_I^2}(m_I^2-m_b^2-m_t^2)\ I^-\left(\frac{m_b^2\ m_t^2}{(m_I^2-m_b^2-m_t^2)^2}\right)\Bigg)\nonumber \\ \nonumber \\
&-&\dfrac{m_b^2}{2}(m_I^2-m_W^2)(2m_t^2\ \eta_U e^{i\alpha_U} + (m_I^2-2m_b^2-m_W^2)\ \eta_D e^{-i\alpha_D}) C_0(0,m_I^2,m_W^2,m_b^2,m_b^2,m_t^2)\nonumber \\ \nonumber \\
&+&\dfrac{m_t^2}{2}(m_I^2-m_W^2)((m_I^2-2m_t^2-m_W^2)\ \eta_U e^{i\alpha_U} + 2m_b^2\ \eta_D e^{-i\alpha_D})  C_0(0,m_I^2,m_W^2,m_t^2,m_t^2,m_b^2)\nonumber \\ \nonumber \\
&-&(m_t^2\ \eta_U e^{i\alpha_U}-m_b^2\ \eta_D  e^{-i\alpha_D})(m_I^2-m_W^2)\left(1-\dfrac{(m_t^2-m_b^2)}{2m_I^2}\log\left(\frac{m_b^2}{m_t^2}\right)\right)\Bigg]\nonumber \\ \nonumber \\
&+&\frac{9}{4}v^2\Bigg[ i (\lambda_4 s_4 - \lambda_5 s_5)\ \left(-2m_W^2 g\left(\dfrac{m_I^2}{m_W^2}\right)+ 2m_I^2\  f\left(\dfrac{m_I^2}{m_W^2}\right) + m_I^2\ I_s(1) + m_W^2\ (1+2g(1))\right)\nonumber \\ \nonumber \\ 
&+&(\lambda_4 c_4 - \lambda_5 c_5)(m_W^2-m_R^2-m_I^2)\ I^+\left(\frac{m_R^2\ m_I^2}{(m_W^2-m_R^2-m_I^2)^2}\right) \nonumber \\ \nonumber \\
&-&(\lambda_4 c_4 - \lambda_5 c_5)\left(-\dfrac{m_W^2m_R^2}{m_I^2}\ I^+\left(\frac{m_I^2}{m_R^2}\right) + \frac{(m_I^2-m_W^2)}{2}\frac{(m_I^2-m_R^2)}{m_I^2}\ \log\left(\frac{m_I^2}{m_R^2}\right)\right) \nonumber \\ \nonumber \\
&-&(m_I^2-m_W^2) (\lambda_4 c_4 - \lambda_5 c_5) \nonumber \\ \nonumber \\
&\times&\Bigg(m_R^2\ C_0(0,m_I^2,m_W^2,m_R^2,m_R^2,m_I^2)\ + m_I^2\ C_0(0,m_I^2,m_W^2,m_I^2,m_I^2,m_R^2)\Bigg)\nonumber \\ \nonumber \\ 
&-&(\lambda_4 c_4 - \lambda_5 c_5)\ (m_I^2-m_W^2)\dfrac{}{}\Bigg]\Bigg\}.
\end{eqnarray}
The remaining loop functions can be written as,
\begin{eqnarray}
I^+(x)&=&\left\lbrace
\begin{array}{ccc}
\sqrt{1-4x}~\ln\left(\dfrac{1-\sqrt{1-4x}}{2\sqrt{x}}\right) & & \text{for }x<\dfrac{1}{4}\\ & & \\
\sqrt{4x-1}~\arcsin\left(\dfrac{\sqrt{4x-1}}{2\sqrt{x}}\right)& & \text{for }x>\dfrac{1}{4}
\end{array}\right.
\end{eqnarray}
\begin{eqnarray}
I^-(x)=\left\lbrace
\begin{array}{ccc}
\sqrt{1-4x}~\left(\ln\left(\dfrac{1-\sqrt{1-4x}}{2\sqrt{x}}\right)+i\pi\right) & & \text{for }x<\dfrac{1}{4}\\ & & \\
-\sqrt{4x-1}~\arccos\left(-\dfrac{1}{2\sqrt{x}}\right)& & \text{for }x>\dfrac{1}{4}
\end{array}\right.
\end{eqnarray}
$g(x)$ is given in Eq.~\ref{mygx} and has the special value $g(1)=\dfrac{\pi}{2\sqrt{3}}$. 

The second form factor is,
\begin{eqnarray}\hspace*{-1.0cm}
\tilde F^{Wg} = \frac{1}{s_W}\Bigg\{
-\frac{3i}{2}|V_{tb}|^2\Bigg[m_t^2\ \eta_U e^{i\alpha_U} C_0(0,m_I^2,m_W^2,m_t^2,m_t^2,m_b^2)\nonumber \\ \nonumber \\
+m_b^2\ \eta_D e^{-i\alpha_D} C_0(0,m_I^2,m_W^2,m_b^2,m_b^2,m_t^2)\Bigg]\frac{}{}\Bigg\}
\end{eqnarray}

\section{\boldmath $H\to gg$}

To validate our  calculation of the one-loop amplitudes we also compute the well known $H\to gg$ \cite{Gunion:1989we,Djouadi:2005gi}. Using the notation
\begin{eqnarray}
{\cal L}(hgg)&=&\frac{\alpha_s}{4\pi v}(F_R^a\ G_{\mu\nu}^A G^{B\mu\nu} + F_R^b\ \tilde{G}_{\mu\nu}^A G^{B\mu\nu})\ H\ \delta^{AB},
\end{eqnarray}
we find agreement with the well known result:
\begin{align}
F_R^a =
I_q\left(\dfrac{m_t^2}{m_H^2}\right)
+\dfrac{3}{4}\dfrac{v^2}{m_H^2}\ &\Bigg[\ \ (\lambda_1+\lambda_2-2\lambda_3)\ I_s\left(\dfrac{m_I^2}{m_H^2}\right) \nonumber \\
&+  (\lambda_1+\lambda_2+2\lambda_3)\  I_s\left(\dfrac{m_R^2}{m_H^2}\right)
+  2\lambda_1 I_s\left(\dfrac{m_{S^\pm}^2}{m_H^2}\right)\Bigg]
\end{align}
and $F_R^b=0$.
In terms of these we have
\begin{eqnarray}
\Gamma(H\to gg)= \dfrac{2}{\pi}\left(\dfrac{\alpha_s}{4\pi v}\right)^2 m_H^3 \left(|F_R^a|^2+|F_R^b|^2\right)
\end{eqnarray}

\bibliography{biblio}

\end{document}